# Mechanical Properties of Graphene Nanoribbons

Ricardo Faccio[a,b(*)], Pablo A. Denis[c], Helena Pardo[a,b], Cecilia Goyenola[a,b] and Álvaro W. Mombrú[a,b]

[a]Crystallography, Solid State and Materials Laboratory (Cryssmat-Lab), DETEMA, Facultad de Química, Universidad de la República, Gral. Flores 2124, P.O. Box 1157, Montevideo, URUGUAY.

[b]Centro Nano*Mat*, Polo Tecnológico de Pando, Facultad de Química, Universidad de la República, Cno. Aparicio Saravia s/n, 91000, Pando, Canelones, URUGUAY.

[c]Computational Nanotechnology, DETEMA, Facultad de Química, Universidad de la República, Gral. Flores 2124, CC 1157, 11800 Montevideo, Uruguay.

(*) corresponding author: rfaccio@fq.edu.uy

## Abstract

Herein, we investigate the structural, electronic and mechanical properties of zigzag graphene nanoribbons upon the presence of stress applying Density Functional Theory within the GGA-PBE approximation. The uniaxial stress is applied along the periodic direction, allowing a unitary deformation in the range of ±0.02%. The mechanical properties show a linear-response within that range while the non-linear dependence is found for higher strain. The most relevant results indicate that Young's modulus is considerable higher than those determined for graphene and carbon nanotubes. The geometrical reconstruction of the C-C bonds at the edges hardness the nanostructure.

Electronic structure features are not sensitive to strain in this linear elastic regime, being an additional promise for the using of carbon nanostructures in nano-electronic devices in the near future.

## I. INTRODUCTION

Quite recently a new carbon nanostructure, called graphene nanoribbon (GNR), has emerged, taking the attention of the scientific community because of its promising use in spintronics. It is manly attributed to the work of Son et al [1,2], who predicted that in-plane electric field, perpendicular to the periodic axis, induces a half-metal state in zigzag nanoribbons (ZGNR). Apart from the interesting dependence of the electronic structure upon an electric field, this is a promising material for future spintronic devices, since it could work as a perfect spin filter. Very recently Campos-Delgado *et al* [3] reported a chemical vapour deposition route (CVD) for the bulk production of long, thin, and highly crystalline graphene ribbons (less than 20-30 μm in length), with widths from 20 to 300 nm and small thicknesses (2 to 40 layers). This experimental advance further increases the expectations for the use of these materials in high-tech devices.

In parallel there is an increased interest in the physical properties of carbon nanostructures in general, due to their outstanding mechanical and electronic properties. Recently, Lee *et al* [4] measured the mechanical properties of a single graphene layer, demonstrating that graphene is the hardest material known, since the elastic modulus reaches a value of 1.0 TPa. Besides, many efforts have been dedicated to study the electronic properties of graphene, because creating a gap could allow the use of graphene in field effect transistors. Many mechanisms have been proposed with that purpose: nano-patterning, creating quantum dots, using multilayers, covalent functionalization [5], doping with heteroatoms such as sulfur [6] and applying mechanical stress [7,8]. In this last case, within linear elasticity theory and a tight-binding approach, Pereira *et al.* [8] observed that strain can generate a bulk spectral gap. However this gap is critical, requiring threshold deformations in excess of 20%, and only along preferred directions with respect to the underlying lattice.

The evidence presented above clearly points that it is important to know how the electronic properties of ZGNR depend on stress, in order to predict its performance in future devices (e.g. gates).

In literature, many representative results concerning the simulation mechanical properties of carbon nanostructures can be found. In particular classical methods have been widely and successfully applied to: polymerized nanotubes [9], nanotube networks [10], 'super' carbon nanotubes [11] and Möbius & twisted graphene nanoribbons [12]. However there are just very few reports related with the study of strain in graphene nanoribbons [13-15] and neither of them report the Young's modulus. The main conclusions from these works indicate that there is no important variation of the electronic properties of zigzag nanoribbons upon stress-strain effects (i.e. energy gaps and local magnetic moments), while there is no information regarding the mechanical properties of this nanostructure.

To the best of our knowledge, in this work we present the first systematic determination of the Young's modulus, Poisson's ratio and calculated Shear modulus for graphene nanoribbons. The paper will be structured as follows. In section I.i we describe the state of the art regarding mechanical properties of carbon nanoribbons and related nanostructures. In section I.ii we briefly review the most relevant features of the electronic structure of ZGNR for the present simulation. In section III, we present and discuss the Young's modulus, Poisson's ratio and shear Stress for different ZGNR.

**I. i. Mechanical properties of carbon nanostructures**

The Young's modulus is a measure of the stiffness of a solid and together with two more elastic parameters defines the mechanical properties of the material: the Young's modulus "E", the shear modulus "G" and the Poisson's ratio "$\upsilon$". In the case of graphene it has more sense to define the in-plane stiffness ($E^{2D}$) instead of the classical 3D Young's modulus ($E^{3D}$), because of the reduced dimensionality of this material. For this reason in graphite the elastic properties can be considered independent of the interlayer distance between graphene, $c_0$=3.35 Å, and the Young's modules can be described as follows:

$$E^{2D} = \frac{1}{A_0}\left(\frac{\partial^2 E_S}{\partial \varepsilon_x^2}\right)_{E_0} = E^{3D} c_0$$ , where $E_S$, $\varepsilon_x$ and $A_0$ corresponds to the total energy, linear strain and equilibrium reference area of the 2D material respectively. The in-plane stiffness of graphite is obtained considering an axial load over graphene. The value obtained in this case is $E^{3D}$=1.02(3) TPa [16]. It allow us to obtain $E^{2D}$= 3.41(9) TPa·Å. This value is almost identical to that obtained experimentally for graphene, $E^{2D}$=3.42(30) TPa·Å [4], using nano indentation with atomic force microscope. This result is in agreement with those reported by Kuddin *et al* [17] and Van Lier *et al* [18]. Using *ab initio* methods they reported Young's modulus of $E^{3D}$= 1.02 TPa·Å [17] and 1.11 TPa·Å [18]. The Poisson's ratio is unambiguously defined in terms of the transversal ratio over the longitudinal variation with a value of $\upsilon$= 0.149. Many representative results for

graphene and nanotubes, based on Reddy *et al* [19] and others, are presented in Table 1 [10,17-30].

The single-walled carbon nanotubes (SWCNT) are an example of a one dimensional system described in terms of 2D property $E^{2D}$, since two parameters must be informed, the tube length (L) and the tube radius (r), in order to gain independence of size effects. Several expressions have been published for their mechanical properties, in terms of multidimensional Young's modulus as: $E^{3D}$, $E^{2D}$, etc [29,30]. The values reported show a wide variation on experimental $E^{nD}$'s values, up to an order of magnitude of difference. This is mainly due to the difficulty in determining the precise structure of nanotubes under study, the presence of defects, chirality, etc. Recently, Wu *et al* [29] used a combined optical characterization of individual SWCNT, coupled with magnetic actuation technique, to measure the Young's modulus of nanotubes with known chirality. The Young's modulus was $E^{3D}$=0.97(16) TPa, assuming a wall thickness of c=3.4 Å corresponding to the interlayer spacing in graphite. No dependence on the nanotube's chiral index within the experimental accuracy was found. This result agrees quite well with theory, in particular with the values reported by Bogár *et al* [30]. Employing an all electron DFT method, they reported $E^{2D}$ for different tubes radius, that ranges from r=1.32 Å to 4.11 Å. They concluded that there is no dependence between the Young's modulus and the chirality of the nanotube.

**I. ii. Electronic and geometrical structure of zigzag nanoribbons**

In graphene nanoribbons, the presence of different types of boundary shapes, called edges, modifies the electronic structure of the material. The major effects are observed at the Fermi level, displaying unusual magnetic and transport features [31]. The zigzag edges (ZGNR) present electronic localized states at the boundaries, corresponding to non-bonding states that appear at the Fermi level as a large peak in the density of states. The non-magnetic solution has many states at the Fermi level, which produces a strong instability that can be resolved by spin polarization or geometrical distortion. Due to the non-bonding character of the zigzag localized edge states, the geometrical reconstruction is unlikely to happen [32] and the spin polarization of the electronic density, establishes an antiferromagnetic arrangement with the opening of a gap, yielding a Slater insulator [33]. The opening of the gap is related with the ZGNR width, since it is a consequence of the interaction between edges. For this reason wider ribbons, with longer distances between opposite edges, recovers the graphene geometry with a gap equal to zero. The tendency observed corresponds to an exponential decay of the energy gaps when increasing the nanoribbon's width (N). Table 2 shows our results for N= 4, 5, 6, 7, 8, 9 and 10 (see Figure 1).

## II. METHODS

The theoretical study of the uniaxial stress on different ZGNR is based on the First Principles – Density Functional Theory [34,35] which we successfully used to study, bulk graphene, thioepoxidated SWCNT, sulfur doped graphene and double wall

CNT [6,36-38]. The simulations are performed using the SIESTA code [39-41] which adopts a linear combination of numerical localized atomic-orbital basis sets for the description of valence electrons and norm-conserving non-local pseudopotentials for the atomic core. The pseudopotentials were constructed using the Trouiller and Martins scheme [42] which describes the interaction between the valence electrons and atomic core. We selected a split-valence double-$\zeta$ basis set with polarization orbitals for all the carbon atoms. The extension of the orbitals is determined by cutoff radii of 4.994 a.u. and 6.254 a.u. for $s$ and $p$ channels respectively, as obtained from an energy shift of 50 meV due to the localization. The total energy was calculated within the Perdew–Burke–Ernzerhof (PBE) form of the generalized gradient approximation GGA xc-potential [43]. The real-space grid used to represent the charge density and wavefunctions was the equivalent of that obtained from a plane-wave cutoff of 230 Ry. The atomic positions were fully relaxed in all the cases using a conjugate-gradient algorithm [44] until all forces were smaller than 10 meV/Å was reached. A Monkhorst Pack grid [45] of 300x2x2 supercell, defined in terms of the actual supercell, was selected to obtain a mesh of 600 k-points in the full Brillouin Zone. All these parameters allow the convergence of the total energy, which corresponds to the antiferromagnetic solution in all the cases.

In order to validate our methodology we calculated the Young's modulus of (5,5) SWCNT, for which the literature shows several results from *ab initio* methods (see Table 1). The smallest unit cell contains a total of 20 carbon atoms. With the purpose to study the dependence on the number of carbon atoms, we simulated the case 40 carbon atoms per unit cell. The Young's modulus obtained are $E^{3D}$=1.03(2) and $E^{3D}$=1.01(3) TPa, for 20 and 40 carbon atoms in the unit cell respectively. The results are consistent within the uncertainty, which was estimated from the variance obtained from the adjustment of the second order fitting of the energy upon unitary deformation. Therefore one can conclude that the results are not affected by the number of supercells used along the periodic direction.

Additionally the results are in good agreement with the reported in the bibliography; see Table 1, in particular with an excellent agreement with those from Bogár *et al* [30]. However there exist some differences in the Young's modulus of graphene obtained by Classical Methods. Force Field approaches seem to underestimate the Young's modulus of graphene by 20 % [10]. In the case of Brenner potentials, it has been demonstrated the strong dependence of $E^{3D}$ on the equilibrium adjustment yield used in the calculation [19]. The Young's modulus changes from 1.11 TPa to 0.7 TPa when the potential is optimized. For this reason the comparison should be done taking into account the methodology involved in the simulation.

Regarding geometry in graphene nanoribbons, we can distinguish two C-C bond orientations: the bond perpendicular to the crystalline periodic direction d(|) and the bond diagonal to the normal direction d(/). The bond distances differ from the inner part of the ribbon (bulk) respecting the atoms at the edge. In the case of bulk C-C distances we found $d(|)_{bulk}$= 1.44 Å and $d(/)_{bulk}$=1.44 Å, while at the edge of the ribbon we found $d(|)_{edge}$= 1.46 Å, and $d(/)_{edge}$=1.43 Å. This result agrees with the tendency observed by Pisani *et al* [33], where the perpendicular bond elongates at the edge, contracting the corresponding diagonal bond at the edge. It promotes an increase of the zigzag C-C-C angle from 120° at the bulk to 121.9° at the edge. This trend is observed for the whole un-stressed studied ribbons.

For all of these reasons, we can unequivocally conclude that our methodology is valid.

The selected ZGNR for simulation correspond to N=4, 5, 6, 7, 8, 9 and 10. Since the code handled was designed for three dimensional materials, we designed special unit cells. All the cells were orthogonal, with the GNR placed in the *ab* plane, and oriented with the periodic direction along the *a* axis, see Figure 1 for the ZGNR N=4 sketch. In order to avoid interference between symmetry images, vacuum regions of 15 Å were added along *b* and *c* directions. In the case of the smallest unit cell, the *a* axis value for every cell is approximately $a_0$= 2.495 Å, with a total number of atoms of 2N+2. With the purpose of increase the number of degrees of freedom in each case, the cells were expanded in four units along the *a* axis ($a=4a_0$), it allows us to multiply by four the number of atoms inside the supercells according to 8N+8. The total number of atoms in each case is: 40, 48, 56, 64, 72, 80 and 88.

## III. RESULTS AND DISCUSSION

The stress-strain curves are obtained applying different stress to the GNR, allowing full atomic relaxation together with full unit cell parameters optimization, until the desired stress tensor is reached. Since we are considering uniaxial strain only, the Voigt tensor has only one non-zero component: $[\sigma_x, \sigma_y, \sigma_z, \sigma_{xy}, \sigma_{xz}, \sigma_{yz}] \Rightarrow [\sigma_x, 0, 0, 0, 0, 0]$. The selected stress components of the Voigt tensor allow us to establish strains in the range of $\varepsilon_x$= ±0.020 for the whole series, which assures a linear stress regime [46,47]. It corresponds to a quadratic dependence of the total energy upon the strain. The most important features of the data treatment are presented in Figure 2 for N=10 ZGNR.

While the second derivative of the total energy is easily obtained, the reference surface is ambiguously defined, with a dependence of the results upon the surface selection. In particular, the problem arises with the selection of the GNR's width, since it is a surface of pruned edges. In our case we have selected two different ways of determine the reference width of GNR: the shortest C-C width ($d_A$) and the longest C-C width ($d_B$). A sketch of these distances is presented in Figure 3(a). It is clear that neither of them are the best selection, and it becomes a problem when we want to compare these results in the N-infinity limit, corresponding to graphene. For this reason all the results are presented, together with the results for graphene. Figure 3(b) shows the variation of the $E^{2D}$ upon the GNR's width N. The same results are presented in Table 3. To check the reliability of the calculations, the case of N=∞ (graphene) was studied. In this case we take a rectangular supercell with 32 carbon atoms. Each periodic crystalline axis were oriented along the zigzag and armchair directions, selecting a *c* value of 20 Å in order to avoid interference between images. The stress was applied along the zigzag axis. The obtained Young's modulus $E^{3D}=E^{2D}/c_0=$ 0.964(9) TPa agrees quite well with early reported values (see Table 1), as well as the Poisson's ratio $\upsilon=$ 0.17, that matches with the one reported by Kudin *et al* [17] $\upsilon=0.149$ and Liu *et al* [48] $\upsilon=0.186$. It is another point that helps us to validate our methodology.

It is important to note that the most differing results correspond $E^{3D}=$ 0.799 TPa. The universal force field seems to overestimate the bulk modulus and to underestimate the basal plane Young's modulus by 20%, in the case of perfect crystalline structures.

The results show $E^{2D}_A$ and $E^{2D}_B$ decreases while N increases, always having a Young's modulus higher than the graphene one. We can argue that ZGNR are harder than graphene. This tendency is the opposite of the case for carbon nanotubes, and the reason can be easily explained in terms of graphene bending. The curvature of CNTs softens the rolled-up graphene sheet because of the lost of overlapping between of the $sp^2$ orbitals, with a pronounced effect for smaller tubes [30]. In the case of GNR the sheet is always plane, with a perfect $sp^2$ overlapping and strong stiffness. In principle, this result would not be expected, but the response could be understood qualitatively in terms of two opposing effects: the curvature of graphene and geometrical edge reconstruction. The higher the curvature the lower the orbital overlap, and hence the lower of the hardness. Furthermore our results indicate that the energy necessary to deform the ribbons (strain energy), expressed as energy per atom, is lower when more carbon atoms are involved, thus fewer atoms hardness of the material. The origin of this effect lies in the geometrical reconstruction of the C-C bonds positioned at the edge. As was mentioned in the introduction, the diagonal C-C distances of GNT at the edges contracts ~ 0.02 Å at the same time that the zigzag C-C-C angle increases ~2°, orientating the stronger C-C diagonal bonds more parallel to the periodic direction of the nanostructure and hardening the bonds. This effect is more evident in the case of thin GNR since there are few C-C bulk bonds, and as the GNR width increases, the bulk bonds prevails diluting the effect of the harder C-C bonds at the edge. In the case of nanotubes the relaxation effect on the edge does not exist, and therefore the curvature effect prevails.

The Poisson's ratio presents a similar tendency to the one observed for the Young's modulus. The results are shown in Figure 4 and Table 4, where the $v_i = -\varepsilon_y^i / \varepsilon_x$ ($i$= A and B) values are presented together with the value for graphene. The tendency between $v$ and N corresponds to a damped oscillation in the case of $v_A$, while the dependency is smoother for the case of $\upsilon_B$. In an extrapolated limit the infinite widths the ratios $\upsilon_i$ are: $\upsilon_A$= 0.18 and $\upsilon_B$= 0.22.

The shear modulus can be obtained using: $G^{3D} = E^{3D}/2(1+\upsilon)$. For graphene we obtained $G^{3D}$= 0.408 TPa. This value agrees with $G^{3D}$=0.384 TPa reported by Reddy *et al* [19], but differs in almost two times with those reported for Sakhaee-Pour [49]. Employing a force field method for finite graphene sheets, with different edge terminations, he obtained G values that range from 0.21 to 0.23 TPa. It is important to note that the corresponding Poisson's ratio reported by Sakhaee-Pour was calculated using E and $\upsilon$. However our results are more similar to those reported for SWCNT [50,51] for which there have been reported $G^{3D}$ values ranging from 0.250 to 0.485 TPa. This is a valid reference for our results, since in this case the mechanical load involves only a single graphene layer. This is the main reason why shear modulus of SWCNT are higher than MWCNT, since in this last case there exist sliding effect between nanotubes that reduces the shear modulus. On the one hand, this discrepancy can be attributed to the different nature of the methods used for the simulation. On the other hand our results were estimated for two independent parameters $E^{2D}$ and $\upsilon$. Regarding the dependence of Shear modulus upon ribbon width, see Figure 5 and Table 4, what we found is a similar dependence to Young's modulus vs. N. This is an expected result since shear modulus expression is dominated by its numerator, corresponding to the Young's modulus, while the denominator remains almost constant, since the Poisson's ratio remains almost constant. Further simulations, including shear deformation, should be done in order to shed more light on this subject.

Regarding electronic structure features of GNR we found no significant dependence of its properties upon strain. These results agree to those early reported [13-15], whereas for the case of ZGNR it has been found a small variation of energy gaps and local magnetic moments, with no variation in the ordering of the occupied-bands. In our case the energy gaps increases in $\delta_{Egap}=0.02$ eV for a positive strain of $\varepsilon= 0.02$, and reduces in $\delta_{Egap}= -0.02$ eV for compressive strain $\varepsilon= -0.02$. These results are valid for all the studied GNR's widths. Similar results are obtained for local magnetic moments at the carbon edges, in all the cases the variation are in the order of $\pm$ 3% for the same strain range studied.

## IV. CONCLUSIONS

In summary, the electronic and mechanical properties of stressed ZGNR were calculated using *ab initio* Density Functional Theory. The proposed models allowed us to obtain the corresponding Young's modulus, shear modulus and Poisson's ratio for ZGNR with different width. In all the cases the GNR present higher constants than graphene, but they approximate to this value when the GNR's width is increased. This effect could be explained in terms of the hardness of the C-C bonds positioned at the edges of the GNR, due to observed geometrical reconstruction. This property could lead to important consequences regarding the structure of the edge of this nanostructure because chemical substitution, the appearance of defects, and chemical doping could soft or stiff the edges. All these possibilities could lead to an important variation of the mechanical properties of GNR, in particular for the case of shorter GNR of low dimensional systems. It would be interesting to simulate la presence of strong donating and strong acceptor groups as functional groups substituting the presence of the single H atoms. Regarding the mechanical properties it has been observed a first order dependency of stress upon strain in the region from ε= -0.02 to ε= +0.02. A non-linear dependence is found for higher strain. Electronic structure features are not sensitive to strain in this linear elastic regime, being an additional promise for the using of carbon nanostructures in nano-electronic devices in the near future.

## ACKNOWLEDGMENTS

The authors thank the PEDECIBA-Química and CSIC -Uruguayan organizations- for financial support.

**FIGURE CAPTIONS**

**Figure 1.-** (a) Graphene nanoribbon with N=4, displaying its smallest unit cell; the arrow shows the periodic direction $\vec{a}$. (b) Spin density map, showing the antiferromagnetic arrangement between opposite edges. (Color on-line)

**Figure 2.-** (a) Normalized total energy versus strain, and (b) the corresponding force versus strain for N=10 ZGNR, indicating a liner stress-strain regime.

**Figure 3.-** (a) The N=5 ZGNR sketching the distances: $d_A$(square-black), $d_B$ (circle-red) and $d_C$ (triangle-blue). (b) The $E^{2D}$'s Young modulus according to the different distances considered in the model which are expressed in terms of the ribbon's width (N) in (c). The horizontal blue line corresponds to the graphene results. (Color on-line)

**Figure 4.-** The $\upsilon$'s dependence upon the GNR's width (N). The horizontal blue line corresponds to graphene results. (Color on-line)

**Figure 5.-** Shear modulus $G^{3D}$ for GNR indicating the estimated value for graphene.

**TABLE CAPTIONS**

**Table 1.-** Representative results for different carbon nanostructures.

**Table 2.-** Energy gaps for different zigzag graphene nanoribbons

**Table 3.-** Final $E^{2D}$'s Young modulus obtained from the different GNR width ($d_i$'s)

**Table 4.-** Poisson's ratio and estimated shear modulus for $d_A$ and $d_B$ models

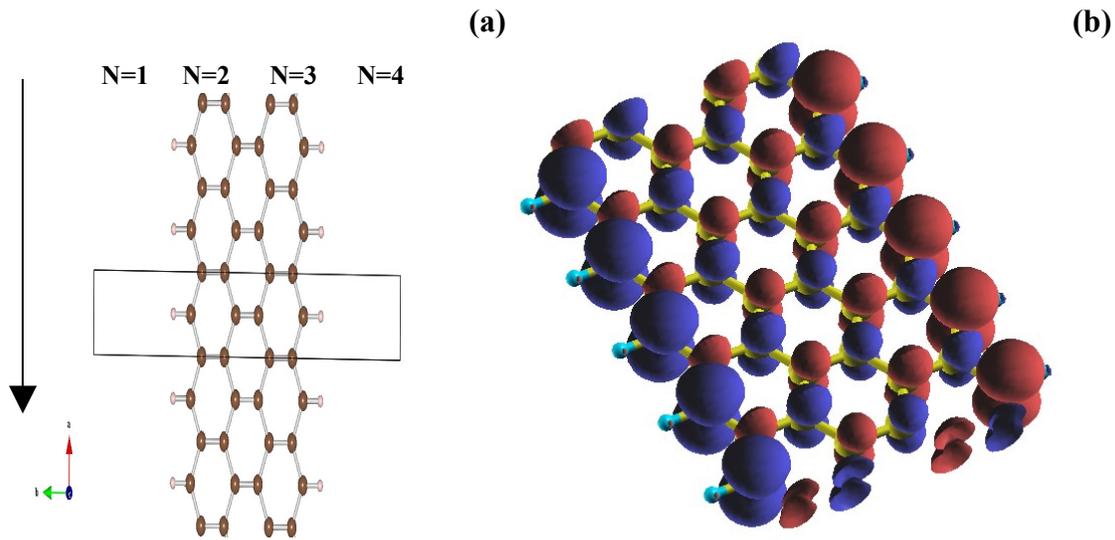

**Figure 1.**

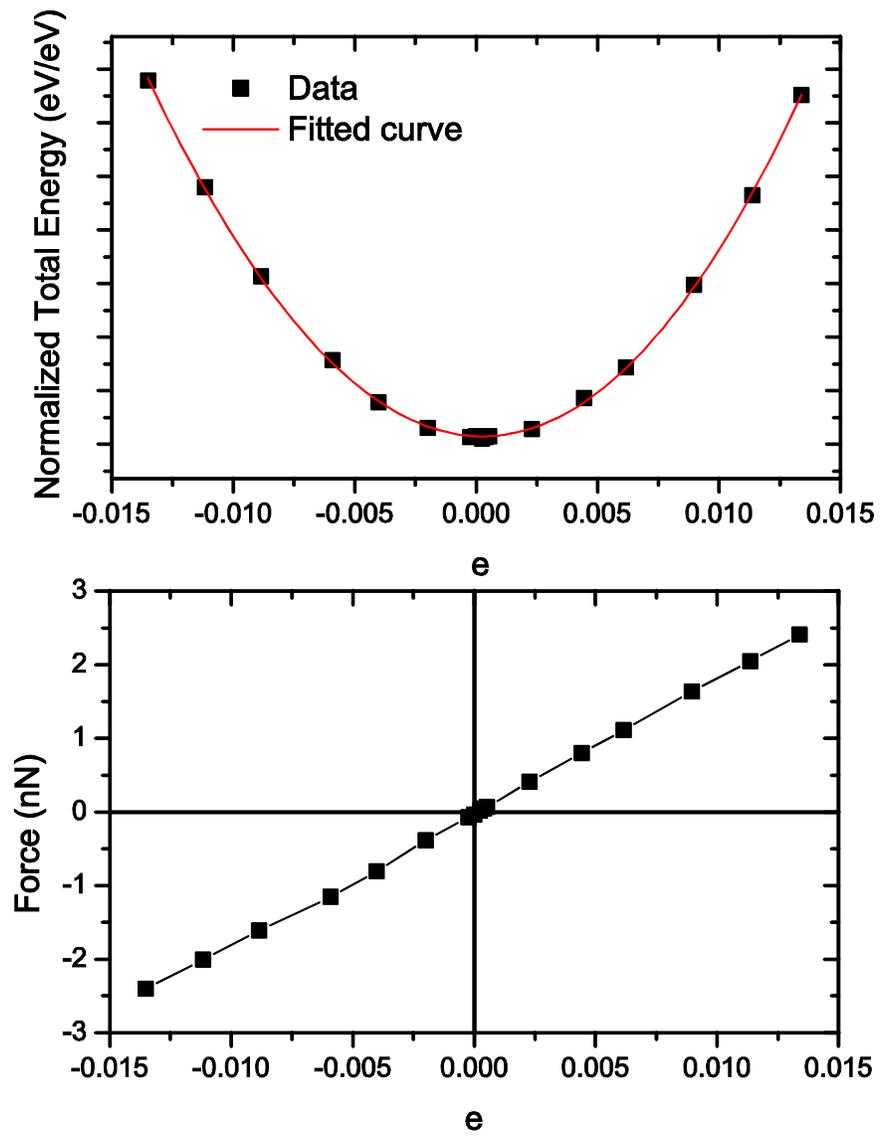

**Figure 2.**

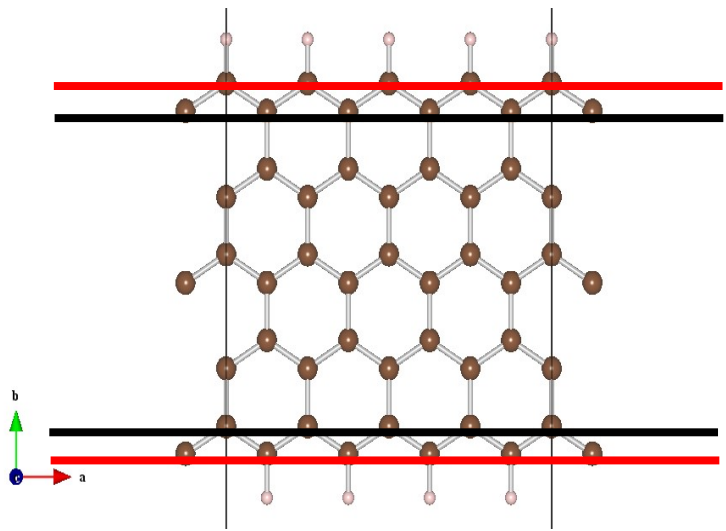

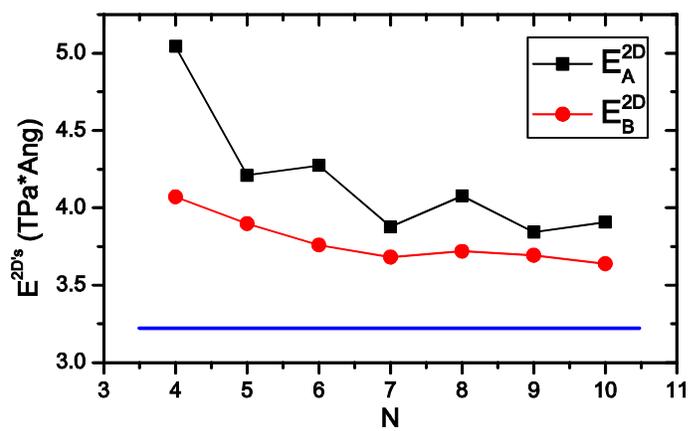

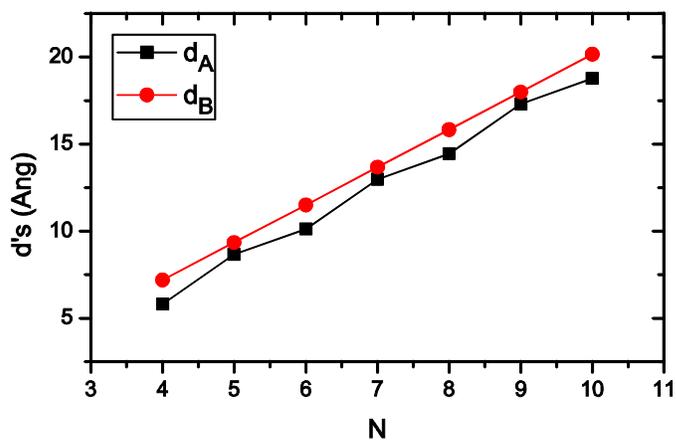

(a)

(b)

**Figure 3.**

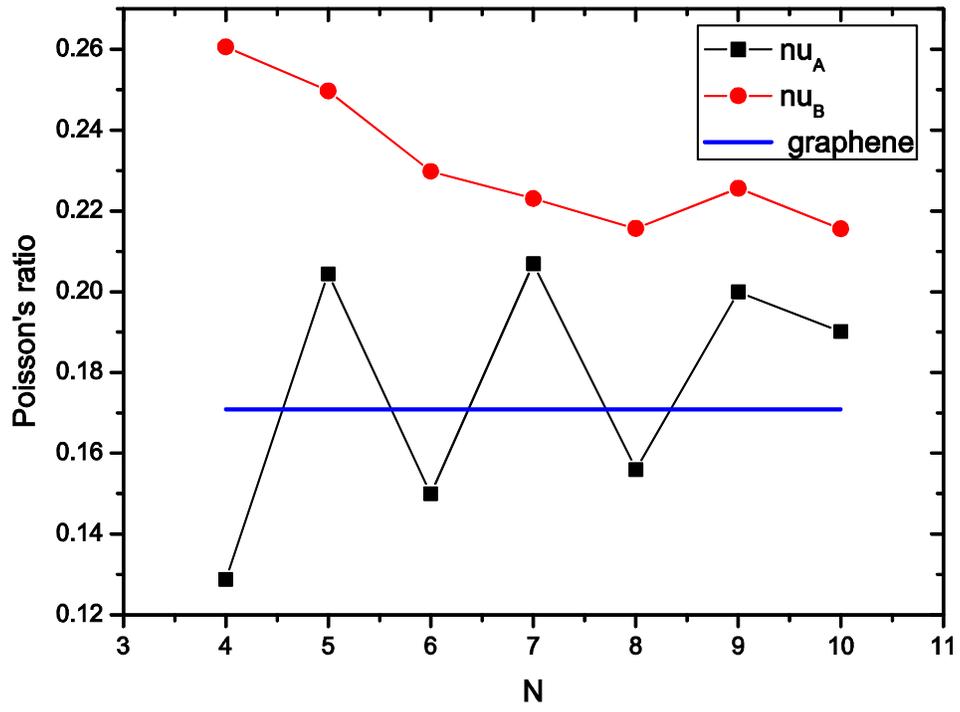

**Figure 4.**

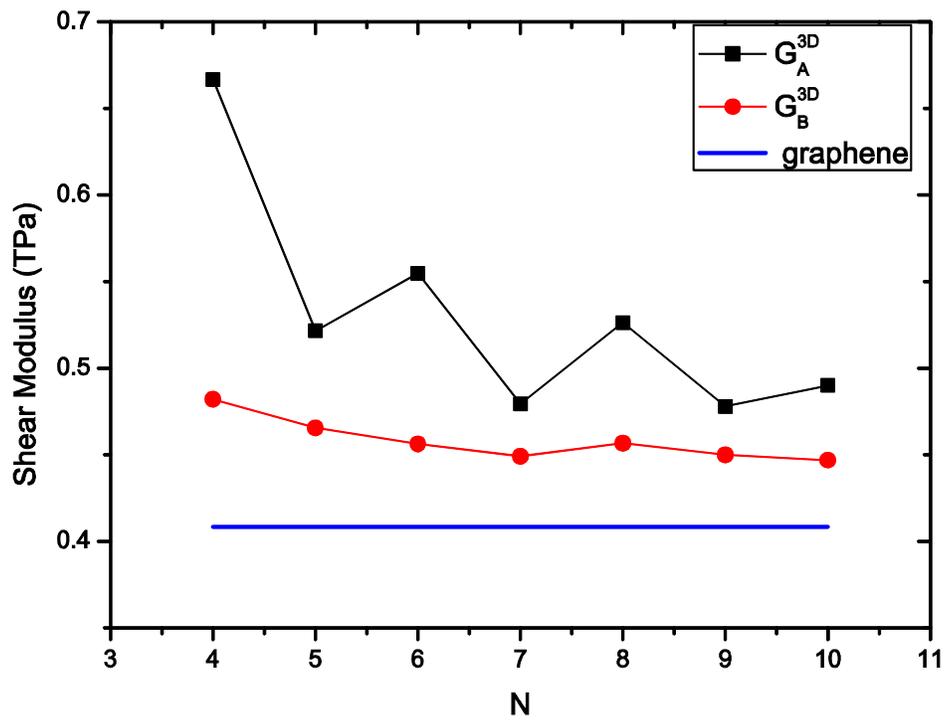

Figure 5.

| Reference | $E^{3D}$ (TPa) | $\upsilon$ | Remarks |
| --- | --- | --- | --- |
| **Graphene** | | | |
| Coluci *et al* [10] | 0.799 | - | Graphene (Force field) |
| Kudin *et al* [17] | 1.02 | 0.149 | Graphene (*ab initio*) |
| Lier *et al* [18] | 1.11 | - | Graphene (*ab initio*) |
| Reddy *et al* [19] | 1.012 | 0.245 | Graphene (Brenner**) |
| Reddy *et al*[19] | 0.669 | 0.416 | Graphene (Brenner*) |
| Arroyo *et al*[20] | 0.694 | 0.412 | Graphene (Brenner) |
| Reddy *et al* [24] | 1.11 | 0.45 | Graphene (Truss model) |
| Present work | 0.96 | 0.17 | Graphene (*ab initio*) |
| **Carbon Nanotubes** | | | |
| Zhang *et al* [21] | 0.694 | - | SWNT (Brenner) |
| Lu *et al* [23] | 0.97 | 0.28 | SWNT (Empirical model) |
| Shen *et al* [25] | 0.213–2.08 | 0.16 | SWNT (MM) |
| Yu *et al* [26] | 0.32–1.47 | - | SWNT (Experiments) |
| Sammalkorpi *et al* [27] | 0.7 | - | SWNT (MD) |
| Yoon *et al* [28] | 1.0 | 0.25 | DWNT (Vibrations) |
| Wu *et al* [29] | 0.81–1.13 | - | SWNT (Experiments) |
| Bogár *et al* [30](***) | 0.8-1.05 | - | SWNTS (*ab initio*) |
| Bogár *et al* [30] | 1.05 | - | SWNT (5,5) - (*ab initio*) |
| Present work | 1.01 | - | SWNT (5,5) - (*ab initio*) |

(*) Minimized potential and (**) Non-minimized Potential

(***) This result was converted to $E^{3D}$ for comparison purposes, using: $E^{3D}=E^{2D} \cdot c_0$.

| Width (N) | Energy Gap (eV) |
|:---:|:---:|
| 4 | 0.63 |
| 5 | 0.59 |
| 6 | 0.54 |
| 7 | 0.50 |
| 8 | 0.46 |
| 9 | 0.43 |
| **10** | 0.40 |

**Table 2.-**

| N | $d_A$ | $d_B$ | $E^{2D}_A$ | $E^{2D}_B$ |
|---|---|---|---|---|
| | Å | | TPa* Å | |
| 4 | 05.80 | 07.19 | 5.04 | 4.07 |
| 5 | 08.66 | 09.35 | 4.21 | 3.90 |
| 6 | 10.12 | 11.51 | 4.27 | 3.76 |
| 7 | 12.98 | 13.68 | 3.88 | 3.68 |
| 8 | 14.45 | 15.83 | 4.08 | 3.72 |
| 9 | 17.30 | 18.00 | 3.84 | 3.69 |
| 10 | 18.77 | 20.16 | 3.91 | 3.64 |
| ∞ | - | | 3.23 | |

**Table 3.-**

| N | $\upsilon_A$ | $\upsilon_B$ | $G_A$ | $G_B$ |
|---|---|---|---|---|
| 4 | 0.129 | 0.261 | 0.667 | 0.482 |
| 5 | 0.204 | 0.250 | 0.522 | 0.466 |
| 6 | 0.150 | 0.230 | 0.555 | 0.456 |
| 7 | 0.207 | 0.223 | 0.480 | 0.449 |
| 8 | 0.156 | 0.216 | 0.526 | 0.457 |
| 9 | 0.200 | 0.226 | 0.478 | 0.450 |
| 10 | 0.190 | 0.216 | 0.490 | 0.447 |
| $\infty$ | 0.179 | | 0.408 | |

**Table 4.-**